\documentclass[twocolumn,prl,showpacs,superscriptaddress]{revtex4}
\usepackage{amsmath}
\usepackage{graphicx}
\usepackage{bm}
\usepackage{ulem}

\begin{document}

\title{An Example of Quantum Anomaly in the Physics of Ultra-Cold Gases}

\author{Maxim Olshanii}
\affiliation{Department of Physics, University of Massachusetts Boston, Boston MA 02125, USA}
\affiliation{Laboratoire de physique des lasers, Institut Galil\'{e}e, Universit\'{e} Paris 13 and CNRS, 
99 avenue J.-B. Cl\'{e}ment, F-93430 Villetaneuse, France}
\author{H\'{e}l\`{e}ne Perrin}
\affiliation{Laboratoire de physique des lasers, Institut Galil\'{e}e, Universit\'{e} Paris 13 and CNRS, 
99 avenue J.-B. Cl\'{e}ment, F-93430 Villetaneuse, France}
\author{Vincent Lorent}
\affiliation{Laboratoire de physique des lasers, Institut Galil\'{e}e, Universit\'{e} Paris 13 and CNRS, 
99 avenue J.-B. Cl\'{e}ment, F-93430 Villetaneuse, France}

\date{\today}

\begin{abstract}
In this article, we propose an experimental scheme for observation of a quantum anomaly---quantum-mechanical symmetry breaking---in a two-dimensional 
harmonically trapped Bose gas. The anomaly manifests itself in a shift of the monopole excitation frequency
away from the value dictated by the Pitaevskii-Rosch dynamical symmetry [L.\ P.\ Pitaevskii and A.\ Rosch, Phys. Rev. A, {\bf 55}, R853 (1997)]. 
While the corresponding classical Gross-Pitaevskii equation and the derived from it hydrodynamic equations do exhibit this symmetry, 
it is---as we show in our paper---violated under quantization.
The resulting frequency shift is of the order of 1\% of the carrier, well in reach for modern experimental techniques. We propose using the 
dipole oscillations as a frequency gauge.
\end{abstract}

\maketitle

{\it Introduction}.--
The current decade is marked by the emerging links between ultracold-atom physics on one hand and cosmology 
and high-energy physics on another.  Examples include kinetics of black holes \cite{garay2000_4643}, 
electron-positron pair production \cite{fedichev2003_011602},
{\it Zitterbewegung} \cite{Vaishnav2008_153002}, and the string theory limits posed on viscosity \cite{johnson2010,thomas_2010}.
However, quantum anomalies are usually considered to be a purely quantum-field-theoretical phenomenon
\cite{treiman1985,donoghue1992}. In this article, we suggest a scheme for observing a quantum 
anomaly in ultracold two-dimensional harmonically trapped Bose gas.

Quantum anomaly---otherwise known as the quantum mechanical symmetry breaking---consists of three ingredients.
First ingredient is an exact symmetry in the classical version of the theory in question.  
Second is a divergence that appears in the straightforward quantum version of the theory. Third
ingredient is a weak violation of the original symmetry that emerges in a regularized version 
of the quantum theory.  

The two-dimensional $\delta$-potential has been long recognized as an example 
of quantum anomaly in elementary quantum mechanics \cite{jackiw1991,holstein1993}.
The classical symmetry of the $\delta_{2}$-potential originates from the following property:
under the scaling transformation ${\bm r} \to \lambda {\bm r}$, the potential transforms in exactly 
the same way as the kinetic energy does. A consequence of this property is the absence of any length scale 
in the corresponding dynamical problem, both before and after a straightforward quantization. 
Next, an analysis of the scattering properties of the $\delta_{2}$-potential \cite{mead1991,nyeo2000} shows a 
divergence in an all Born orders of the scattering amplitude starting from the second. 
Finally, the subsequent regularization \cite{mead1991,nyeo2000} leads to 
the appearance of the new length scale $a_{2D}$. The original symmetry becomes broken, and a quantum anomaly emerges.

Similar anomaly in the case of the $1/r^2$ potential was discussed in Ref.\ \cite{camblong2001_220402}, along with various regularization methods
\cite{camblong_2000_1590,coon2002_513}.

In Ref.\ \cite{pitaevskii1997}, Pitaevskii and Rosch predicted a dynamical symmetry that appears  
in the classical field theory (Gross-Pitaevskii equation) of the $\delta_{2}$-interacting two-dimensional 
harmonically trapped Bose gas. 
This symmetry is a direct consequence of the scaling symmetry of the $\delta_{2}$-potential, described above. 
The consequences of this symmetry are (a) absence of the amplitude dependence of the main frequency (isochronicity)
and (b) absence of higher overtones in the time dependence (monochromaticity) of the monopole oscillations of the moment of inertia.
Both properties were demonstrated experimentally \cite{chevy2001_250402}, along with an anomalously slow damping, 
for the case of a very elongated Bose-Einstein condensate; its Gross-Pitaevskii equation
coincides with the one for a two-dimensional condensate.  

In this article we address the question of {\it whether the Pitaevskii-Rosch symmetry survives quantization}. 

In the fully quantized unitary gas, the analogous symmetry has been shown to remain unbroken \cite{castin2004_407}. 
The quantum correction to the frequency of the monopole excitation in an elongated condensate \cite{chevy2001_250402} 
was computed in Ref.~\cite{pitaevskii1998_4541}

{\it The zero-temperature equation of state of the two-dimensional Bose gas}.--
The low-density zero-temperature quantum field theory (QFT) expression for the chemical potential of the two-dimensional Bose gas is well known 
\cite{popov1983,mora2003}: it reads
\begin{eqnarray}
\mu(n) \stackrel{\mbox{\scriptsize{QFT}}}{=}\frac{4\pi\hbar^2}{m} \chi(\pi e^{2\gamma+1} n a_{2D}^2) \, n
\quad,
\label{mu_0}
\end{eqnarray}
where $a_{2D}$ is the two-dimensional scattering length, $\gamma = 0.5772\ldots$ is the Euler's constant,
$n$ is the two-dimensional density, and $\chi(z) = \frac{1}{-W_{-1}(-z)}$. For a given two-dimensional
two-body interaction potential $V({\bf r})$, its scattering length $a_{2D}$
is defined as the radius of a hard disk whose zero-energy $s$-wave scattering amplitude equals the 
one for the potential $V$.
Here, $W_{-1}(z)$ is the minus-first branch of the Lambert's W-function \cite{Wm1}. For the small values 
of the gas parameter $n a_{2D}^2$, the factor $\chi$ is a logarithmically slow function of the 
density: $\chi(z) \stackrel{z \to 0}{\approx} 1/\ln(1/z) + {\cal O}(\ln(\ln(1/z))/\ln(1/z)^2 )$.
The expression (\ref{mu_0}) is an inverse of a more conventional formula 
$n(\mu) = (m \mu/4\pi\hbar^2) \ln(4\hbar^2/e^{2\gamma+1} m\mu a_{2D}^2)$,
shown to be the leading term in an expansion in powers of $1/\ln(4\hbar^2/e^{2\gamma+1} m\mu a_{2D}^2)$
\cite{mora2009_180404}.

In the case of three-dimensional short-range-interacting atoms tightly confined to a two-dimensional plane by a one-dimensional harmonic 
potential, the two-dimensional scattering length can be expressed through 
the three-dimensional scattering length $a_{3D}$ and the confinement 
size $\tilde{a}_{z} = \sqrt{2\hbar/(m\omega_{z})}$ as  
$a_{2D} = C_{2D} \tilde{a}_{z} \exp[-(\sqrt{\pi}/2)(\tilde{a}_{z}/a_{3D})]$,
where $C_{2D} = 1.4795\ldots$ \cite{petrov2001,pricoupenko2007}. The chemical potential becomes 
$\mu(n) \stackrel{\mbox{\scriptsize{QFT}}}{=} \bar{g}_{2D} \, n\, (\chi(\sigma(n) e^{-\frac{1}{\epsilon}})/\epsilon)$
where $\bar{g}_{2D} = 4\pi\epsilon\hbar^2/m$ is the ``bare'' two-dimensional coupling constant,
$
\epsilon = a_{3D}/\sqrt{\pi}\tilde{a}_{z}
$
is the {\it small parameter governing the proximity to the classical limit},
and $\sigma(n) = \pi e^{2\gamma+1} (C_{2D})^2 n \tilde{a}_{z}^2$.
In the limit $\epsilon \ll  \min\left(1,\, 1/|\ln(\sigma(n))|\right)$,  
the factor $\chi$ approaches the density-independent constant $\epsilon$, and  
the chemical potential converges to the prediction of the classical field theory (CFT), otherwise 
known as the Gross-Pitaevskii equation; there, the chemical potential reads  
\begin{eqnarray}
\mu(n)
\stackrel{\mbox{\scriptsize{CFT}}}{=} \bar{g}_{2D} \, n
\quad.
\label{mu_2}
\end{eqnarray}
%
%

{\it Hydrodynamic equations}.--
The zero-temperature hydrodynamic (HD) equations for a two-dimensional
harmonically trapped Bose gas read
\begin{eqnarray}
&&
\frac{\partial}{\partial t}n + {\bm \nabla}_{\bf r}(n {\bm v})
=
0
\label{continuity_1}
\\
&&
\frac{\partial}{\partial t} {\bm v} + ({\bm v}\cdot {\bm \nabla}_{\bf r}) {\bm v}
=
-(1/m) {\bm \nabla}_{\bf r}[\mu(n) + V_{\mbox{\scriptsize{HO}}}({\bf r})]
\,,
\label{Euler_1}
\end{eqnarray}
where $m$ is the atomic mass, $n = n({\bm r},\,t)$ is the atomic density, ${\bm v} = {\bm v}({\bm r},\,t)$ is the atomic velocity,
$V_{\mbox{\scriptsize{HO}}}({\bf r})=m \omega^{2} r^{2}/2$ is the trapping potential energy per
particle, $\mu(n)$ is the chemical potential, and ${\bm r} = x {\bm e}_{x} + y {\bm e}_{y}$. Assuming that the gas is a Bose condensate with no vortices, 
the velocity field can be assumed to be irrotational:  ${\bf v} =\hbar {\bm \nabla} \Phi/m$, where $\Phi({\bf r})$ is the potential of the velocity field.    
Under this assumption, the HD equations (\ref{continuity_1},\ref{Euler_1}) can be rewritten in a Hamiltonian form,
$
\frac{\partial}{\partial t} n({\bf r},\,t) 
=
\frac{i}{\hbar} [H,\, n({\bf r},\,t)]_{\mbox{\scriptsize HD}}
$
,
$
\frac{\partial}{\partial t} \Phi({\bf r},\,t) 
=
\frac{i}{\hbar} [H,\, \Phi({\bf r},\,t)]_{\mbox{\scriptsize HD}}
$.
The Hamiltonian is represented by a sum of two parts, the first being in turn a sum of the kinetic and interaction 
energies and the second being the trapping energy: $H = H_{0} + H_{\mbox{\scriptsize{HO}}}$. 
Here  
$
H_{0} 
 = \int \!d^2{\bf r} \, \left\{ 
                         \frac{\hbar^2}{2m} n ({\bf \nabla} \Phi)^2 + \varepsilon(n)
                       \right\}
$,
$
H_{\mbox{\scriptsize{HO}}} = \int \!d^2{\bf r} \, V_{\mbox{\scriptsize{HO}}}({\bf r}) 
$,
the ``hydrodynamic commutator'' $[\cdot,\,\cdot]_{\mbox{\scriptsize HD}}$
is given by the Poisson brackets with respect to the $(n,\,\Phi)$ canonical pair,
\begin{eqnarray*}
&&
[A(n({\bf r},\,t),\Phi({\bf r},\,t)),\,B(n({\bf r}',\,t),\Phi({\bf r}',\,t)) ]_{\mbox{\scriptsize HD}}
=
-\frac{i}{\hbar} \times
\\
&&\,
                 \left[
                 \left(  \frac{\partial}{\partial n}   A(n({\bf r},\,t),\Phi({\bf r},\,t))  \right) 
                 \left(  \frac{\partial}{\partial \Phi}B(n({\bf r},\,t),\Phi({\bf r},\,t))  \right)
                 -
                 \right.
\\
&&\,
                 \left.
                 \left(  \frac{\partial}{\partial \Phi}A(n({\bf r},\,t),\Phi({\bf r},\,t))  \right) 
                 \left(  \frac{\partial}{\partial n}   B(n({\bf r},\,t),\Phi({\bf r},\,t))  \right)
                 \right]
\times
\\
&&\qquad
\delta_{2}({\bf r}-{\bf r}')
\quad,
\end{eqnarray*}
and $\varepsilon(n) = \int_{0}^{n} \! dn' \, \mu(n')$ is the microscopic energy density. 

{\it The Pitaevskii-Rosch symmetry and the quantum anomaly at the HD level}.--
Introducing the generator 
of the scaling transformations,
$
Q = \frac{1}{\hbar} \int \!d^2{\bf r} \, n ({\bf r}\cdot{\bf \nabla} \Phi) 
$
(see Ref.\ \cite{pitaevskii1997} for example),
one obtains the following set of commutation relations:
\begin{eqnarray}
[Q,\,H_{0}]_{\mbox{\scriptsize HD}} \stackrel{\mbox{\scriptsize QFT}}{=} 2 i H_{0} + i a_{2D} \frac{\partial}{\partial a_{2D}} H_{0}
\quad,
\label{com_Q_H0__Q}
\end{eqnarray}
$
[Q,\,H_{\mbox{\scriptsize{HO}}}]_{\mbox{\scriptsize HD}} \stackrel{\mbox{\scriptsize QFT}}{=} - 2 i H_{\mbox{\scriptsize{HO}}}
$,
$
[H_{\mbox{\scriptsize{HO}}},\,H_{0}]_{\mbox{\scriptsize HD}} \stackrel{\mbox{\scriptsize QFT}}{=}  i \omega^2 Q  
$.   
Notice that the classical field theory chemical potential (\ref{mu_2}) does not depend on the two-dimensional 
scattering length $a_{2D}$. Then, the commutator (\ref{com_Q_H0__Q}) becomes 
\begin{eqnarray}
&&
[Q,\,H_{0}]_{\mbox{\scriptsize HD}} \stackrel{\mbox{\scriptsize CFT}}{=} 2 i H_{0} 
\quad.
\label{com_Q_H0__C}
\end{eqnarray}
In this case, the observables $H_{0}$, $H_{\mbox{\scriptsize{HO}}}$, and $Q$ form a closed 
three-dimensional algebra, identical to the one discovered by Pitaevskii and Rosch \cite{pitaevskii1997}
at the Gross-Pitaevskii equation level.

However, the more accurate quantum field theory prediction for the chemical potential (\ref{mu_0}) depends explicitly 
on $a_{2D}$. According to the Eqn.\ \ref{com_Q_H0__Q}, the classical commutation relation (\ref{com_Q_H0__C}) becomes corrected by 
$
i a_{2D} \frac{\partial}{\partial a_{2D}} H_{0}
$.
Since the correction term is not generally expected to be a function of three original 
members of the algebra, the algebra opens; such an opening constitutes a {\it quantum anomaly}.

{\it Castin-Dum-Kagan-Surkov-Shlyapnikov equations}.--
If the factor $\chi$ in the chemical potential expression 
(\ref{mu_0}) was a constant, the HD equations 
(\ref{continuity_1},\ref{Euler_1}) could be solved via 
the Castin-Dum-Kagan-Surkov-Shlyapnikov (CDKSS) scaling ansatz \cite{castin1996,kagan1996_R1753}. Note however,
that $\chi$ is a very slow function of the density, and thus the 
scaling ansatz should approximately hold.
Consider the following ansatz:
$
n({\bm r},\,t) = \frac{1}{\lambda(t)^2} n_{0} [1 - (r/(\lambda(t)R))^2]
$,
$
{\bm v}({\bm r},\,t) = {\bm e}_{r} r \frac{\dot{\lambda}(t)}{\lambda(t)} 
$,
where $\tilde{R}_{TF} = \sqrt{2g_{2D}(n_{0}) n_{0}/m \omega^2}$ 
is the steady-state HD curvature of the density distribution
(that also corresponds to the Thomas-Fermi radius of a gas with a density-independent 
coupling constant fixed to $g_{2D}(n_{0})$),
$g_{2D}(n) \equiv (\partial \mu/\partial n)$ is an effective density-dependent 
coupling constant, 
$n_{0}$ is the steady-state peak density; the scaling parameter 
$\lambda$ obeys the CDKSS equation
\begin{eqnarray}
&&
\ddot{\lambda} = \frac{\omega^2 u(\lambda)}{\lambda^3} - \omega^2 \lambda
\label{lambda_eqn}
\end{eqnarray}
with $u(\lambda) = g_{2D}(n_{0}/\lambda^2)/g_{2D}(n_{0})$.
It can be shown that the above ansatz solves the HD equations 
(\ref{continuity_1},\ref{Euler_1}) almost everywhere, with the exception of an exponentially 
narrow ring close to the edge of the cloud. 

For example, let us define the thickness $\Delta R(\delta)$ of a ring of positions ${\bf r}$ such that the effective coupling $g_{2D}(n(r))$  differs 
from that in the center by a relative correction $\delta$
or more, but the density is still greater than zero: $|g_{2D}(n(r))/g_{2D}(n_{0}) -1| > \delta$ and $r < R_{TF}$. 
(Here, $R_{TF}$ is the ``true'' Thomas-Fermi radius defined by $V_{\mbox{\scriptsize HO}}(R_{TF})=\mu(n_{0})$.)
Then one can show that in the limit 
$\epsilon \to 0, \, \sigma(n) = \mbox{const}$, the leading behavior of the thickness $\Delta R(\delta)$ 
is $\Delta R(\delta) = (1/2) \exp(-\delta/\epsilon)\, R_{TF}$. 
Likewise, one can show that the curvature radius 
$\tilde{R}_{TF}$ is exponentially close to the Thomas-Fermi radius of the cloud: 
$\tilde{R}_{TF} \approx R_{TF}$. 

{\it Anomalous frequency shift of the monopole frequency}.--
Linearization of the equation (\ref{lambda_eqn}) for small excitation amplitudes readily gives
the frequency of small oscillations,
$
\Omega = \Omega_{0} ( 1 + \delta(\mbox{AR}_{\lambda}\!=\!0))
$,
where \cite{W_property} 
\begin{eqnarray}
\delta(\mbox{AR}_{\lambda}\!=\!0)) 
= \frac{1}{4} \, \frac{\chi(\sigma(n_{0}) e^{-\frac{1}{\epsilon}})}{(1-\chi(\sigma(n_{0}) e^{-\frac{1}{\epsilon}}))^2}
\quad.
\label{delta_CD}
\end{eqnarray}
The deviation of the monopole frequency $\Omega$ from the 
classical field theory prediction $\Omega_{0} \equiv \Omega \Large|_{\epsilon \ll  \min\left(1,\, 1/|\ln(\sigma(n))|\right)}  = 2\omega$ is a manifestation 
of the quantum anomaly.

Here and below,
$\mbox{AR}_{\lambda} = (\lambda_{max}-\lambda_{min})/(\lambda_{max}+\lambda_{min}) = (\lambda_{max}^2-1)/(\lambda_{max}^2+1)$
is the aspect ratio for the monopole oscillations. 

For $\epsilon \ll  \min\left(1,\, \frac{1}{|\ln(\sigma(n_{0}))|}\right)$,
the relative anomalous correction converges to
\begin{eqnarray}
\delta(\mbox{AR}_{\lambda}\!=\!0)) \approx \frac{1}{4\sqrt{\pi}} \frac{a_{3D}}{\tilde{a}_{z}}
\quad.
\label{delta_small_a3D}
\end{eqnarray}
In particular, this estimate shows that the effect of the anomaly can be enhanced using 
a Feshbach resonance. 

\begin{figure}
\includegraphics[scale=.7]{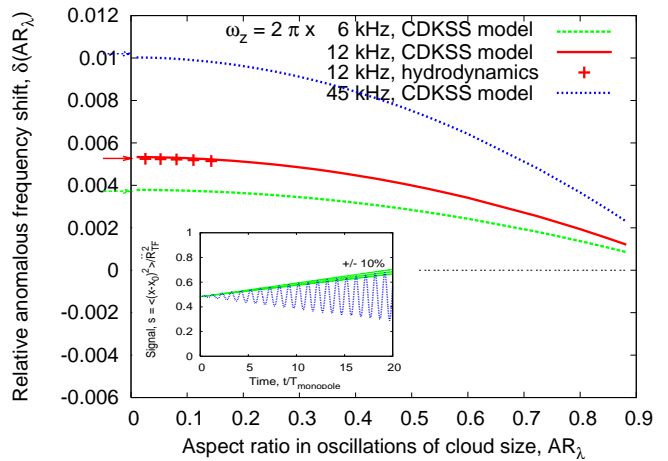}
\caption
{
\label{fig:delta_omega} (color online). 
Relative anomalous shift of the monopole frequency, $\delta(\mbox{AR}_{\lambda}) \equiv (\Omega-\Omega_{0})/\Omega_{0}$ as a function of the amplitude of the excitation, for different 
values of the transverse confinement frequency $\omega_{z}$.
Solid lines: prediction of the modified Castin-Dum-Kagan-Surkov-Shlyapnikov (CDKSS) model (\ref{lambda_eqn}). Crosses: 
prediction of the HD equations (\ref{continuity_1},\ref{Euler_1}). Arrows: 
the $a_{3D} \ll \tilde{a}_{z}$ limit (\ref{delta_small_a3D}) of the CDKSS model prediction (\ref{delta_CD})
for the zero-amplitude shift. The rest of the parameters corresponds to $N=10^{3}$ rubidium 87 atoms 
($a_{3D} = 5.2\, \mbox{nm}$) confined longitudinally in a frequency $\omega$ harmonic trap,
where $\omega = 2\pi \times 41\, \mbox{Hz}$ for the $\omega_{z} = 2\pi \times 45\,\mbox{kHz}$ data and $\omega = 2\pi \times 40\, \mbox{Hz}$
for the rest of the data. The steady-state peak density $n_{0}$ entering the  CDKSS model was derived from the number of atoms $N$ using the Euler equation (\ref{Euler_1}).
An possible experimental setup corresponding to the $45\,\mbox{kHz}$ curve is described in detail in the concluding section of this article.
Insert shows the monopole vs.\ dipole beat signal for the optimal choice of the initial trap shift (see main text).
Parameters correspond to the $45\, \mbox{kHz}$ curve taken at the aspect ratio
$AR_{\lambda} = 0.486$.
We also show the initial linear part of the slow envelope,
$(\lambda_{max}^2/6 + ((\lambda_{max}^4 - 1)/(12 \lambda_{max}^2)) \, \delta(\mbox{AR}_{\lambda}=0.486) \,  \Omega_{0} t$ plotted
with a $\pm 10\%$ uncertainty in the value of the anomalous correction $\delta(\mbox{AR}_{\lambda}) = 0.00760$. 
The CDKSS model was used for calculations.
}
\end{figure}
Furthermore, a numerical analysis shows that under quantization, the monopole frequency 
becomes amplitude-dependent.
Fig.~\ref{fig:delta_omega} shows the corresponding prediction 
of the CDKSS equation (\ref{lambda_eqn}) compared 
to both the results of the full HD treatment (\ref{continuity_1},\ref{Euler_1}) and the limiting values 
(\ref{delta_small_a3D}). 

{\it A possible experimental scheme for detecting the anomalous frequency shift}.--
Consider the following excitation scheme. Initially, the cloud is prepared in the ground state of 
a frequency $\omega_{\mbox{\scriptsize{init.}}}$ trap at a position $x=0,\,y=0$. At $t=0$, the trap is simultaneously relaxed to 
a {\it lower} frequency $\omega$ and shifted to a new position $x=x_{0},\,y=0$. This initial condition will induce 
a superposition of a monopole oscillation that starts from the lowest in cloud size turning point 
and a dipole oscillation---whose frequency $\omega$ does not depend on either interaction strength or the oscillation 
amplitude---that starts from the left(right) turning point for $x_{0}>0$($x_{0}<0$).

What we suggest is to measure  
the spatial mean of the square of the horizontal displacement with respect to the new trap center, 
$s(t) = \int dx\,dy\,  (x-x_{0})^2 n(x,\,y,\,t)$.
After a lengthy but straightforward calculation it can be shown that this observable will evolve in time as 
$
s(t)/\tilde{R}_{TF}^2 = P + A \cos[\Omega_{0}t] + B \cos[\Omega_{0}(1+\delta(\mbox{AR}_{\lambda}))t + \phi]
$,
where
$
P = (\lambda_{max}^4 + 1)/12 \lambda_{max}^2 + \eta^2/2
$,
$
A = \eta^2/2
$,
$
B = (\lambda_{max}^4 - 1)/12 \lambda_{max}^2
$, 
$
\phi = \pi
$,
$\eta = x_{0}/\tilde{R}_{TF}$, and $\tilde{R}_{TF}$ is the effective Thomas-Fermi radius for the {\it final}, frequency $\omega$ trap
(see a definition above).
Choosing 
$
\eta = \sqrt{\frac{\lambda_{max}^4 - 1}{6\lambda_{max}^2}}
$
leads to $A = B$. 
It is easy to see that in this case, at $t=0$  
the beats between the dipole and monopole oscillations have a node, with the first 
crest reached in $1/(2\delta(\mbox{AR}_{\lambda}))$ monopole periods.
In terms of the initial trap, the optimal shift of the trap can be expressed as 
$
x_{0} = \frac{1}{\sqrt{6}} \sqrt{\left(\frac{\omega_{\mbox{\scriptsize{init.}}}}{\omega}\right)^2 - 1} \,\, \tilde{R}_{TF,\,\mbox{\scriptsize{init.}}}
$,
where $\tilde{R}_{TF,\,\mbox{\scriptsize{init.}}} = \sqrt{\omega/\omega_{\mbox{\scriptsize{init.}}}} \tilde{R}_{TF}$ 
is the Thomas-Fermi radius for the {\it initial}, frequency $\omega_{\mbox{\scriptsize{init.}}}$ trap.
Note as well that the right turning point for the oscillations of the scaling parameter can be expressed as 
$\lambda_{max} = \sqrt{\omega_{\mbox{\scriptsize{init.}}}/\omega}$. In our excitation scheme, the monopole oscillations start from the left turning point that in turn corresponds to the lowest 
cloud size. Insert of Fig.~\ref{fig:delta_omega} presents an example of the projected beat signal for a typical set of parameters. 

The present in any realistic trap anharmonicity and anisotropy may complicate the matters. 
For the anharmonicity, one can show that for a quartic correction of a form $m \omega^2 r^2/2 \to (m \omega^2 r^2/2)\, (1 + \alpha r^2/2 a_{0}^2)$ 
($a_{0} = \sqrt{\hbar/m\omega}$ being the size of the ground state), the relative (to the monopole frequency) shift of the monopole frequency  
is $\delta_{\mbox{\scriptsize anharm.}} = \frac{12}{11}\alpha\sqrt{N\epsilon}$, to the leading order in both $\alpha$ and the amplitude. 
The doubled dipole frequency---that serves as a reference---is unshifted in that order. 
On the other hand, the anisotropy leads to a splitting of the dipole frequency with the monopole frequency situated in between the two resulting 
frequencies.   
If one chooses the doubled X-dipole frequency $2\omega_x$ as a reference,  
the relative to this frequency monopole shift will be given, to the lowest order in both
anisotropy $\omega_x-\omega_y$ and the amplitude,
by  $\delta_{\mbox{\scriptsize anisotr.}} = (\omega_{x}-\omega_{y})/2\omega_{x,y}$, 
where $\omega_{x,y}$ is any of the two frequencies. 
Both the anharmonicity ($\delta_{\mbox{\scriptsize anharm.}}$) and anisotropy ($\delta_{\mbox{\scriptsize anisotr.}}$) shifts 
must be kept below 1\% to allow for observation of the quantum anomaly.

Vortex-antivortex pair creation has been suggested as the dominant mechanism for damping of the two-dimensional monopole oscillations  \cite{fedichev2003_011602}, all the conventional
channels being suppressed due to the Pitaevskii-Rosch symmetry. Here, the amplitude of the oscillations decays                                        
as $\lambda_{max}(t) = \lambda_{max}(t)/(1+t/\tau)^{1/10}$, where
$
\tau = \frac{737280}{127}  \left( \frac{\omega_{\mbox{\scriptsize{init.}}}}{\omega} \right)^5 \left(\frac{\lambda_{min}(0)}{\lambda_{max}(0)}\right)^{10}   
\frac{ \ln(4 m\bar{g}_{2D} N / \sqrt{\pi} \hbar^2) }{  (m\bar{g}_{2D}/\hbar^2)^{9/4} \sqrt{N} }  
$. Here, it is assumed that the amplitude of oscillations is large (i.e.\ $\lambda_{max} \gg 1$) and the gas is dilute (i.e.\ $a_{3D} \ll \tilde{a}_{z}$). 

Let us now propose a concrete example of a trap suited for an observation of the quantum anomaly. The numbers proposed are those of 
rubidium 87 in its $5S_{1/2},F=2,m_F=2$ ground state. A blue detuned standing wave (wavelength $532\,\mbox{nm}$, power $P$, waist $w_{0}$) 
crosses, at a right angle, the symmetry axis $z$ of a quadrupolar magnetic field;  the magnetic field gradients are given by $b'$ in the $xy$ plane and $-2b'$ along $z$. 
This results in a series of pan-cake traps: there, the magnetic field is responsible for the weak horizontal trapping, while the standing 
wave gives a strong vertical confinement. Atoms could be loaded in one of the nodes of the standing wave, at a distance $z=d$ 
from the trap center; there the magnetic field is bounded from below by $B_0 = 2b'd$. By construction, this trap should be isotropic 
in the horizontal plane. Furthermore, the residual anisotropy defects could be further compensated by adding small magnetic 
gradients in the horizontal plane. For $b' = 167$\,G/cm, $d=400\,\mu\mbox{m}$, 
%
$P=4\,\mbox{W}$, 
%
and $w_{0}=500\,\mu\mbox{m}$, 
the oscillation frequencies are 
%
$41\,\mbox{Hz} \times 41\,\mbox{Hz} \times 45\,\mbox{kHz}$, 
%
and the magnetic field minimum $B_{0} = 13.4\,\mbox{G}$ well prevents 
any Majorana losses. The photon scattering rate is as low as $2.6\times 10^{-3}\,\mbox{s}^{-1}$, corresponding to 30,000 monopole 
mode oscillations. For $N=10^4$ atoms (ten times the number used in Fig.\ \ref{fig:delta_omega}), 
the anharmonic parameter is bounded from above by $\alpha = -2.1\times 10^{-6}$. The predicted  
Thomas-Fermi radius of $15\,\mu\mbox{m}$ is much smaller than both $d$ and $w_{0}$;   
accordingly, the anharmonic shift has a negligibly small value of $\delta_{\mbox{\scriptsize anharm.}} = 5\times 10^{-5}$. The chemical potential 
of $\mu/h=1.7\,\mbox{kHz}$ is well below the transverse frequency; this ensures that the trap is well in the two-dimensional regime. 
Finally, taking oscillation aspect ratio of $\mbox{AR}_{\lambda} = 0.486$ as an example, one obtains the damping time $\tau$ which is $1576$ times longer than the period 
of the monopole oscillations.  
Overall, these figures bring our proposal 
within reach of modern experimental technology.

{\it Summary and outlook}.--
In this article, we propose an experimental scheme for observation of a quantum anomaly in a two-dimensional harmonically trapped Bose gas. 
The effect consists of a shift of the monopole excitation frequency away from the value dictated by the Pitaevskii-Rosch dynamical symmetry \cite{pitaevskii1997}. 
The shift we predict is only of the order of 1\% of the base frequency. To detect it, we propose using the dipole oscillations 
as a reference frequency.

\begin{acknowledgments}
We are grateful to Eric Cornell, Steven Jackson, and Felix Werner for enlightening discussions on the subject. 
Laboratoire de physique des lasers is UMR 7538 of CNRS and Paris 13 University. 
LPL is member of the Institut Francilien de Recherche sur les Atomes Froids (IFRAF).
This work was supported by grants from the Office of Naval Research ({\it N00014-06-1-0455}) 
and the National Science Foundation ({\it PHY-0621703} and {\it PHY-0754942}).
\end{acknowledgments}

%


\begin{thebibliography}{27}
\bibitem[{Wm1()}]{Wm1}
\bibinfo{note}{$W_{-1}(z)$ is the real $W < -1/e$ solution of $z= W e^{W}$ in
  the range $-1 < z < 0$.}
\bibitem[{W_p()}]{W_property}
\bibinfo{note}{Note that we have used the property $z \frac{d}{d z} \chi(z) =
  \chi(z)^2/(1 - \chi(z))$ that follows from the properties of the Lambert's
  W-function.}
\end{thebibliography}

%
\end{document}